\documentclass[a4paper,11pt]{article}
\usepackage{pos}

\title{Calculational Techniques in Particle Theory}

\author[a,b]{Lorenzo Tancredi}

\affiliation[a]{Physik Department, 
  Technische Universit\"at Mu\"nchen, James-Franck-Straße 1, 85748 Garching, Germany}

\affiliation[b]{Rudolf Peierls Centre for Theoretical Physics,
  Clarendon Laboratory, Parks Road, Oxford OX1 3PU, UK}

\emailAdd{lorenzo.tancredi@tum.de}

\abstract{In this contribution, I review some of the latest advances in calculational
techniques in theoretical particle physics. I focus, in particular, on their application to
the calculation of
highly non-trivial scattering processes, which are relevant for precision phenomenology studies 
at the Large Hadron Collider at CERN.
}

\FullConference{%
  *** The European Physical Society Conference on High Energy Physics (EPS-HEP2021), ***\\
  *** 26-30 July 2021 ***\\
  *** Online conference, jointly organized by Universität Hamburg and the research center DESY ***\\
  Preprint Number: TUM-HEP-1370/21
}

\begin{document}
\maketitle

\section{Introduction, SM and Precision Physics}
The discovery of the Higgs Boson at the CERN LHC in 2012~\cite{ATLAS:2012yve,CMS:2012qbp} has been a milestone in particle physics,
providing firm ground for the impressive theoretical edifice of the Standard Model of Particle Physics.
Despite the fact that, with the Higgs discovery, the full matter content of the Standard Model has now been observed,
there are still many unanswered questions about its theoretical foundations. 
One of the most pressing issues is  the origin and structure of the Yukawa interactions
between the Higgs boson and the fermions (quarks and leptons). 
Among the other things, the way the Higgs couples to fermions, in particular the patterns of its
interactions with the first family of quarks and leptons, determines the masses of those particles
and, ultimately, it is the reason why chemistry has allowed for the existence of life as we know it.

Until recently, the only indication that we had of the Higgs coupling to fermions was indirect, namely
through the observation of its production in gluon fusion and decay to photons; in the Standard Model, both are
in fact loop-induced processes and happen through the coupling
of the Higgs to the heavy top quark.
A real breakthrough took place in 2018, when both ATLAS and CMS observed, for the first time,
the production of a $t \bar{t}$ pair in association with a Higgs boson~\cite{ATLAS:2018mme,CMS:2018uxb}, $t\bar{t}H$.
This was the first direct observation of the coupling of the Higgs boson with top quarks. 
After this result, more have followed,
in particular confirming that the Higgs couples, roughly as expected, to the third family of quarks and leptons,
see for example recent results in~\cite{ATLAS:2020qdt}.

But let us keep our focus on $t\bar{t}H$, which provides a perfect example of the 
challenges that the experimental results obtained at the LHC are posing to the theory community.
Already in their first measurements in 2018, ATLAS and CMS could get a good estimate of the relevant 
Yukawa coupling, with an overall precision around $\sim \mathcal{O}(15-20 \%)$.
This  measurement was performed by analysing a small fraction of the 
full expected data-set that will be produced at the LHC. Indeed, compared the total amount of data
that will be collected starting with Run 3 and continuing
through the High-Luminosity phase (HL-LHC), only $5\%$ of the total collisions
expected have already been delivered. A naive estimation would suggest that the statistical error
on these measurements
could go down of a factor of 6 at HL-LHC, bringing it down to the few percent level.

Keeping up with such  precision on the theoretical side is very challenging.
The theoretical modelling of high-energetic collisions at the LHC is in fact notoriously difficult,
due to the interplay of physics at very different energy scales. 
Here, the existence of (collinear) factorisation theorems in Quantum Chromodynamics (QCD) allows us to separate
non-perturbative long-range (low-energy) physics from perturbative short-range (high-energy) physics effect
in a consistent way, such that corrections to the factorised picture are expected to be of the order
of $\mathcal{O}(\Lambda_{QCD}/Q)$, where $\Lambda_{QCD}$ is the typical non-perturbative QCD scale and 
$Q$ is the scale of the process considered. 
At the scales of LHC
collisions such corrections can be expected to be at the percent level,
which justifies our ambition to try and describe theoretically scattering processes in QCD with a comparable precision.
Special care should be used, anyways, as factorisation theorems are only rigorously proven for a very limited number
of processes with relatively simple kinematics. 

Even assuming the correctness of the factorised picture for complex final states as $t\bar{t} H$, 
a lot of effort has then to be
devoted on the precise determination of the non-perturbative building blocks that make up this picture, most notably the 
parton distribution functions; in addition, the precise simulation of realist icprocesses, including parton showers, 
hadronisation, fragmentation, detector response etc, must also be understood. 
Only by dedicating an equal effort to understanding all these pieces of the puzzle, 
can we dedicate our attention to study high-energetic (hard) collisions
of fundamental partons with percent level precision.
In what follows, I will report on many new results and fascinating new methods developed to the aim
of improving our understanding of fixed-order calculations for complex LHC processes.

\section{Fixed-order calculations for LHC processes}
Quantum Field Theory (QFT) provides us with a set of well defined rules which allow us, under certain assumptions,
to improve consistently our prediction for a given scattering process.
Schematically, in QFT we evaluate cross-sections (and more in general differential observables)
by first computing so-called scattering amplitudes $\mathcal{A}$, which provide the probability amplitudes for
the partonic scattering to take place. Very schematically, 
an observable $\mathcal{O}$ can be computed as
 $$\sigma_\mathcal{O} \sim \int \left| \mathcal{A} \right|^2 d {\rm PS}\,, $$
where $d {\rm PS}$ symbolises the integration measure over the relevant phase-space that defines
the observable considered. 

The scattering amplitudes $\mathcal{A}$ are complex functions of the kinematics and of the
spins and polarisations of the relevant particles, and can be computed perturbatively in
QFT by an expansion in Feynman diagrams. 
Importantly, in order to increase the perturbative order, we typically need to compute Feynman
diagrams with more loops and higher multiplicities, or, in jargon, with more  legs.
While these calculations can be extremely difficult, the many results obtained in the past years
have revealed  unexpected regularities in the final analytical expressions for the scattering amplitudes, 
which in turn have fuelled the study of their mathematical properties from a general perspective.  
In what follows, I will review some of the most recent developments in this field.

\section{New ideas for Scattering Amplitudes}
The standard text-book approach to compute scattering amplitudes goes through the enumeration and
evaluation of the relevant Feynman diagrams that contribute to the scattering process at the
perturbative ordered considered. While straightforward in principle, 
computing Feynman diagrams can quickly become extremely cumbersome.
First of all, there is clearly an obvious combinatorial reason why we \emph{rightfully} expect 
scattering amplitudes to increase in complexity with the number of loops and  legs.
Nevertheless, it has  been observed over and over again in all scattering amplitudes calculations, 
that the apparent complexity  
which one has to deal with in the intermediate stages of a calculation, is often orders of magnitude
worse than the real complexity of the final, physical results.
The main reason why this happens, is the way we perform perturbative calculations in QFT:
by insisting in using \emph{off-shell} Feynman diagrams, which carry multiple redundancies 
due to the virtual particles running in the loops and to gauge freedom, we hide the simplicity
of on-shell, gauge invariant scattering amplitudes.

In order to review the most recent developments in this field, 
I find it useful to separate the types of problems that one encounters when
performing these calculations, into three separate categories: 
\begin{enumerate}
\item Obtaining the integrand, namely the expression which needs to be integrated.
\item Reducing the integrand to master integrals.
\item Computing these master integrals either numerically or analytically.
\end{enumerate}

The issues which stem from the three points above are clearly connected and a 
strict separation among them is of course impossible. I will nevertheless use this
simplifying categorisation to summarise the recent advances in the different directions.

\subsection{Obtaining the integrand}
As we discussed above, Feynman diagrams provide us with a bullet-proof technique 
which allows us to obtain the \emph{integrand} for any scattering amplitude 
in any (perturbative) Quantum Field Theory.
This approach is plagued by a quickly growing combinatorial complexity
and in recent years a lot of effort has been devoted to 
understand how to construct the integrand avoiding a blind generation and manipulation
of all Feynman diagrams.
By generalising on-shell techniques valid at tree-level~\cite{Britto:2004ap,Britto:2005fq} and one-loop~\cite{Bern:1994zx,Bern:1994cg,Ossola:2006us,Ellis:2011cr},
many ideas have been developed to try and avoid Feynman diagrams entirely also at higher loops,
and build integrands only from \emph{on-shell}, \emph{gauge invariant} data~\cite{Badger:2013gxa,Badger:2015lda,Dunbar:2016gjb,Bourjaily:2020qca}.
In the same spirit, but from a different point of view, interesting results have also recently been obtained on 
extending recursion techniques (based on Feynman diagrams!) to construct two-loop integrands efficiently
starting from lower loop amplitudes~\cite{Pozzorini:2020hkx,Lang:2020nnl,Lang:2021hnw}.
These developments are very promising, and they have made it possible to 
compute various non-trivial amplitudes, in particular in the very special all-plus configuration~\cite{Dunbar:2016aux,Badger:2019djh}.
Nevertheless, none of these techniques is still mature enough to 
provide a workable alternative to Feynman diagrams for generic helicity configurations and in
non-supersymmetric theories.

So while it does not seem likely that we will be able to get rid of Feynman diagrams for QCD calculations
in the very near future, an interesting question is how we can improve the way we manipulate 
the Feynman
diagrams, in order to obtain the sought for integrand in the most
efficient and compact form possible.
The main goal of these manipulations is to resolve the Dirac algebra
and to express the Feynman diagrams in terms of loop-independent Lorentz structures, which multiply 
loop-dependent \emph{scalar Feynman integrals}. The latter are simpler to compute and we
will discuss various techniques to do this in the next two sections.

One standard approach to extract scalar Feynman integrals from scattering amplitudes
in terms of Feynman diagrams, is the so-called \emph{projector method}. Schematically,
this works by 
 decomposing the scattering amplitude, tentatively non-perturbatively, 
 in terms of so-called  \emph{tensors} and invariant \emph{form factors} $F_j$
\begin{equation}
\mathcal{A} = \sum_{j=1}^N F_j T_j\,. \label{eq:tens}
\end{equation}
With a slight abuse of language, we call  ``tensors'' $T_j$ are all Lorentz covariant structures that can be built and 
which are consistent with the symmetries of the problem. The $T_j$ include the relevant polarisation
vectors / spinors that contract the Lorentz indices and transform therefore only under the 
corresponding Little Group.
Once we have  the decomposition in eq.~\eqref{eq:tens}, we can derive projector operators
which single out the corresponding form factors. Again schematically, we define the projector $P_k$ as
\begin{equation}
P_k = \sum_{j=1}^N c_j^{(k)} T_j^\dagger\,, \quad P_k \cdot \mathcal{A} = F_k\,,
\end{equation}
where the action $P_k \cdot \mathcal{A}$ is defined by summing over all spins and helicities of the external particles.
This construction is usually done in the so-called conventional dimensional regularisation scheme (CDR),
ie treating all momenta, internal and external, as $d$-dimensional.
While the method was successfully applied to many $2 \to 1$ and $2 \to 2$  calculations up to two and three loops,
it turns out to scale very badly with the increase of the number of external particles and tends to 
become impractical already for $2 \to 3$ processes, see for example~\cite{Boels:2018nrr}.

This complexity stems from the desire of using this method in CDR.
Nevertheless, it was recently realised that if one is interested in computing the ``helicity amplitudes'' 
for the process in
the 't Hooft-Veltman regularisation scheme (tHV) this method can be simplified extremely~\cite{Peraro:2019cjj,Peraro:2020sfm}.
In fact, one can exploit the fact that many of the tensors $T_j$ in eq.~\eqref{eq:tens} become 
linearly dependent when the dimensionality of the external states is equal to four, in order to 
block-diagonalise the
projector matrix and define a subset of physical tensors whose number matches one-to-one that
of the different helicity amplitudes. As it should be, the number of   helicity amplitudes provides then
 an upper bound to the number of the independent tensors required.
Recently, these ideas have been applied successfully to simplify the calculation of the three-loop corrections to $2 \to 2$ 
scattering in QCD~\cite{Caola:2020dfu,Caola:2021rqz}.
A different approach was also developed, where explicit representations for the polarisation vectors and spinors
are used in order to single out the different helicity configurations from the scattering 
amplitudes~\cite{Chen:2019wyb}
Both approaches have been successfully applied to various $2 \to 3$ scattering processes~\cite{Chawdhry:2021mkw,Chawdhry:2020for,Agarwal:2021vdh}.

A different approach to extract integrands in QCD and which can be applied also
on a Feynman diagram representation of the scattering amplitude
is  integrand 
decomposition~\cite{Mastrolia:2011pr,Mastrolia:2012an,Mastrolia:2016dhn,Badger:2012dv}.
Mixing integrand reduction with unitarity based techniques, has also been a very
successful strategy~\cite{Abreu:2019odu,Badger:2018gip,Abreu:2020xvt},
and has allowed to obtain very non-trivial results for a large number of $2 \to 3$ scattering amplitudes~\cite{Abreu:2018zmy,Abreu:2020cwb,Abreu:2021oya,Abreu:2021asb,
Badger:2018enw,Badger:2021ega,Badger:2021imn,Badger:2021nhg}. 
More on the other
crucial developments that have made these calculations possible will be discussed in the next paragraph.

\subsection{Reducing the integrand to master integrals}
Obtaining the integrand for a complicated scattering amplitude in QCD (or in general in the Standard Model)
is only the very first step towards its calculation.
Typically, after the application of any of the techniques described above,
integrands for two or three-loop problems in realistic theories are expressed in terms of a large number of
scalar integrals, which can go from tens of thousands up to many millions.
Clearly, it is out of question to compute each of these integrals on its own.
Luckily, the vast majority of these integrals are not linearly independent, and can instead
be expressed as linear combinations of a (relatively) small number of so-called \emph{master integrals}.
While there are various ways to achieve this decomposition, the algorithmic use 
of \emph{integration-by-parts identities} (IBPs)~\cite{Tkachov:1981wb,Chetyrkin:1981qh}, 
through the by-now famous Laporta algorithm~\cite{Laporta:2000dsw},
has definitely stand out as one of the most powerful and versatile techniques.
IBPs have been applied to a multitude of problems in QCD, 
from two-point functions up to five loops~\cite{Baikov:2016tgj,Herzog:2017ohr},
to five point functions at two loops, which I will say more about later.

Unfortunately, since the Laporta algorithm maps the problem of reduction to master integrals to
the symbolic solution a large system of linear equations, its blind application to solve complex 
multiloop problems
can quickly become unfeasible even on the largest computers, due to the swell of the intermediate
algebraic expressions. 
While it is fair to say that the impressive increase of computational resources available
to theoretical particle physicists has played an
important role in making many outstanding calculations finally possible, the past years
have witnessed the emergence of revolutionising  ideas that have
lead to a breakthrough in the field.
A first important realisation was the fact that
out of the thousands or millions of IBPs produced in the standard Laporta approach, only
a fraction is really relevant for the reduction of the physical scattering amplitudes.
This happens for two reasons. First of all, because many of the identities are simply not
independent and can be thrown away. This was known for a long time but was, to the best of my knowledge,
first fully exploited only recenly in the public code Kira~\cite{Maierhofer:2017gsa}.
Secondly, one can devise algorithms, based on algebraic geometry, in order to
``slice'' through the space of IBP identities and select among those, only the ones that are so-called
\emph{unitarity compatible}~\cite{Schabinger:2011dz,Ita:2015tya,Kosower:2018obg}. This is a way to say, that we would like to select only those
identities that never introduce integrals with spurious double propagators, which do not
appear in scattering amplitudes, except in very special degenerate configurations.
A lot of work has been done in this direction, leading to important developments
for the simplifications of the systems of integration by parts identities.

In parallel to this, a  large impact has come from the introduction of methods based on
Finite Fields reconstruction.
Indeed, Finite Fields have been used in algebra systems for decades to simplify computationally
expensive operations
as the computation of the great common divisor (gcd) of two polynomials.
The breakthrough in particle physics has come from the realisation that, due to the simplicity of the
final physical results, one can use Finite
Fields reconstruction techniques in order to perform the chain of operations that  express
the finite remainder of complicated scattering amplitudes in terms of independent building blocks,
without ever having to manipulate complicated intermediate expressions~\cite{vonManteuffel:2014ixa,Peraro:2016wsq}.
It turns out, in fact, that by repeating the numerical evaluation of the scattering amplitudes 
many times over finite fields,
one can uniquely reconstruct the analytic form of the rational functions that appear in the final result. 
We stress that the strength of this approach resides in the fact that
the final physical result is often orders of magnitudes simpler than the intermediate steps of the calculation.
In the past few years this technology has been adopted
and improved upon by many groups, both in public~\cite{Peraro:2019svx,Klappert:2019emp,Smirnov:2019qkx,Klappert:2020nbg} and private codes.
\newline

Together with Finite Field reconstruction techniques, also worth of mention is the introduction of 
new algorithms for the decomposition of multivariate rational functions into partial fractions~\cite{Abreu:2019odu,Boehm:2020ijp,Heller:2021qkz}.
Indeed, for a long time partial fraction decomposition has been the standard approach to
simplify complicated rational functions showing up into QFT calculations.
Until few years ago, nevertheless, the advent of
larger computers and better gcd algorithms, had initiated a shift in the particle physics community
to prefer more economic gcd-based representations for the final results. 
Only very recently it has been rediscovered
that the rational functions which show up in scattering amplitudes
 are very poorly represented when put under great common divisor, also due to the presence
 of high poles in their denominators. Take for 
 example the
very compact function
$$R(a,b,c,d) = \frac{1}{(a-b)^6}-\frac{1}{(c-d)^5}\,.$$
By forcing it into a gcd representation, we obtain a much more complicated expression
\begin{align*}
&R(a,b,c,d)  = - \\
& \frac{a^6-6 a^5 b+15 a^4 b^2-20 a^3 b^3+15 a^2 b^4-6 a b^5+b^6-c^5+5 c^4 d-10 c^3 d^2+10 c^2 d^3-5 c d^4+d^5}{(a-b)^6 (c-d)^5}\,.
\end{align*}

Inverting the gcd is in general
 non-trivial, as partial fraction decompositions are not unique when many variables are involved and,
in particular, different choices for the ordering of the variables with respect to which the decomposition is done, 
can lead to various issues as, 
for example, the appearance of undesired spurious poles.
The new algorithms that I hinted to above, 
which are based on well known mathematics~\cite{Leinartas:1978pf,Raichev:2012pf}, 
make it possible to avoid the appearance of spurious poles and,
 to obtain compact and
numerically very stable representations for complicated scattering amplitudes.
Notably, in this way extremely compact representations have been obtained also for
non-leading color amplitudes which involve also
complicated non-planar Feynman integrals~\cite{Boehm:2020ijp,Agarwal:2021vdh,Badger:2021imn}.

\subsection{Computing the master integrals}
In the previous two paragraphs, we have discussed about two 
major issues connected to the algebraic complexity of scattering amplitudes:
how to manipulate the large expressions that emerge 
from multiloop / multileg Feynman integrals and how to efficiently decompose them in
terms of independent master integrals.
In our discussion, we have neglected a very non-trivial aspect of the whole story, 
namely the computation of the master integrals themselves in terms of special functions.
Usually, it is indeed only when the finite remainder of a
scattering amplitude in $d=4$ dimensions 
 is expressed in terms of well defined special functions,
that their simplicity becomes manifest.

Integration is notably a very non-trivial operation and an impressive effort has been devoted 
to the developments of efficient
integration techniques for multiloop Feynman integrals. Given their complexity, the first obvious 
question one has to answer is  whether it makes sense to attempt to evaluate Feynman integrals 
analytically, or if a flexible numerical approach are
bound to be more efficient in the long run. Many advances have happened in both directions and, while it would be impossible to
summarise all papers which are worth being mentioned, I will do my best to point out a couple of results 
which are,
in my very personal opinion, particularly interesting.

\subsubsection*{Analytical techniques}

A lot of progress has been achieved in analytical integration methods
after the realisation that Feynman integrals in dimensional regularisation 
are often naturally  expressed in terms of special classes of functions,
that go under the general name of (Chen) iterated integrals~\cite{Chen:1977oja}.
\newline
Before going into the details of the results obtained, it is important to clarify what do we mean when we say
that a result is ``analytic''. Usually, when we talk about an analytic result we imagine
it to be written in terms of elementary ``known functions'', for example exponentials, logarithms, etc.
As a matter of fact, except for some notable exceptions,
Feynman integrals cannot be expressed in terms of elementary functions, and instead 
more general ``special
functions'' are required, some of which might not have a name yet.
The main problem is then to classify which types of functions are
relevant and to study their properties. 
If we define a new class of functions, I think it is fair to say that a result written written
in terms of them 
can be defined to be analytic as long as:
\begin{enumerate}
\item one has full analytic control on the functional relations
among them.
\item one has full control over their numerical evaluation, through for example series expansions.
\end{enumerate}

An especially celebrated class of special functions which fulfils these requirements and which has found
large applicability in the context of Feynman integrals calculation, are the so-called
multiple polylogarithms (MPLs). They are defined as iterated integrals over linear rational
kernels with single poles
\begin{align}
G(a_1,...,a_n;x) = \int_0^x \frac{dy}{y-a_1} G(a_2,...,a_n;y) \;, \quad G(x) =1 \,, \quad G(\underbrace{0,...,0}_n;x) = \frac{1}{n!} \log^n(x)\,,
\end{align}
see ~\cite{Remiddi:1999ew,Vollinga:2004sn} and references therein.

A revolution in direct integration methods, based for example on the well known Feynman-Schwinger
parametrisation, has happened with the establishment of the criterion of linear reducibility~\cite{Brown:2008um} 
for Feynman
integrals and the development of algorithms to exploit it to compute Feynman integrals in terms of MPLs~\cite{Panzer:2014caa}. Many important
calculations have been performed thanks to this insight, recent examples are
various  form factor integrals up to four loops~\cite{vonManteuffel:2015gxa} and $2 \to 2$ scattering amplitudes 
with masses~\cite{Bonetti:2020hqh}.

Equally impressive analytic results have been obtained by the use of the 
by now renown method of differential equations~\cite{Kotikov:1990kg,Remiddi:1997ny,Gehrmann:1999as},
augmented by the choice of a canonical basis of integrals with unit leading singularities,
as first proposed in~\cite{Henn:2013pwa}. This has made it possible to obtain analytical results in terms of
special classes of Chen iterated
integrals for all master integrals relevant to compute five-point two-loop massless scattering amplitudes.
The relevant special functions have been dubbed ``pentagon functions''~\cite{Chicherin:2020oor}.
More recently, also the first results for the corresponding pentagon functions for five-point functions with one off-shell leg been obtained~\cite{Badger:2021nhg,Chicherin:2021dyp}.
For these classes of problems, also results in terms of multiple polylogarithms have been obtained~\cite{Papadopoulos:2015jft,Canko:2020ylt}, using a variant of the differential equations
method~\cite{Papadopoulos:2014lla}.
I consider these results some of the latest successes of the analytic approach. Both criteria 
enumerated above (ie control over functional relations and numerical evaluation) 
are met by these results, which 
allowed various group to use them to obtain compact and stable 
expressions for all massless $2 \to 3$
processes, as discussed above.
Very importantly, thanks to the newly obtained results for these scattering amplitudes,
the first phenomenological studies of $2 \to 3$ processes in NNLO QCD have started to appear~\cite{Chawdhry:2019bji,Kallweit:2020gcp,Czakon:2021mjy,Chawdhry:2021hkp}.

Before closing the discussion on the analytic methods, it is important to mention the
efforts that are being made to extend our understanding of special functions beyond 
multiple polylogarithms. This line of research has been fuelled by a new geometrical understanding
of iterated integrals of rational functions defined on increasingly complicated Riemann
surfaces, the first example of which are the so-called elliptic polylogarithms. This line of research has sparked a lot of interest from different parts of the community, which would be impossible to summarise in few lines. Some example of recent work in this direct can be found in~\cite{Brown:2013hda,Bloch:2016izu,Adams:2016xah,Ablinger:2017bjx,Remiddi:2017har,Broedel:2017kkb,Klemm:2019dbm} and references therein.

\subsubsection*{(Semi-)Numerical techniques}
Finally, progress has been made also in the context of new \emph{numerical methods} for the calculation of 
multiloop, multileg Feynman integrals.
Clearly, numerical methods have an advantage on analytical ones, as they do not require a
study and classifications of the special functions that can appear in the result.
On the other hand, devising general numerical methods for the calculation of Feynman integrals 
is very non trivial, first  of all because Feynman integrals are divergent both in the UV and the IR.
This requires to write general procedures to isolate the poles and regularise the integrals.
Even when this is done, a numerical evaluation in a physical region in Minkowski kinematics requires to
find a proper contour deformation to avoid all the poles on the real integration domain. These poles
are the origin of the discontinuities of the scattering amplitudes and it is crucial to pick the right integration
contour in order to have well-defined results on the physical Riemann sheet.
Worth of mention is definitely the impressive progress in direct numerical integration of Feynman integrals
using the sector decomposition method~\cite{Borowka:2017idc}, which has made it possible to compute various two-loop 
scattering amplitudes for $2 \to 2$ processes mediated by massive quarks~\cite{Borowka:2016ypz,Chen:2020gae}.

More recently, also a different semi-numerical approach has become increasingly popular
to solve complicated multiloop Feynman integrals.
The idea goes back to the well-known Frobenius method for the numerical solution of differential equations
as series expansions
close to regular singular points. In fact, one can use the (analytical) differential equations satisfied
by the Feynman integrals, and use them to obtain fast converging 
series expansions for the integrals, which can in turn be used
for their numerical evaluation~\cite{Moriello:2019yhu,Hidding:2020ytt}. 
This method has been successfully  applied, in various forms,
to some non-trivial $2 \to 2$ and $2 \to 3$ two-loop Feynman integrals~\cite{Frellesvig:2019byn,Abreu:2020jxa}.

An interesting question is then, with respect to which variable does it make sense to differentiate
in order to get differential equations that are easy to solve numerically and automatically, with as few
user input as possible.
A possible answer has recently been provided in~\cite{Liu:2017jxz}, where it has been proposed to differentiate
with respect to auxiliary complex masses, say generically 
$m_{aux}$, suitably placed on all or only some of the propagators.
This method has two advantages: first, one can obtain all boundary conditions at a value of the 
auxiliary masses going to infinity, from the algorithmic procedure of the large mass expansion. 
Second, since one integrates along the imaginary axis, one is usually far from singularities that are confined
on the real axis and one needs to perform fewer steps in the numerical solution of the differential equations
in order to transport the result from the boundary condition to the physical values of $m_{aux} \to 0$.
As it is easy to imagine, there are some subtleties in how one recovers the limit $m_{aux} \to 0$, as in general
Feynman integrals develop IR singularities when masses go to zero.
Modulo these subtleties, this method can be very powerful and it has been applied already to the calculation
of some non-trivial scattering amplitudes, see for example~\cite{Bronnum-Hansen:2020mzk,Bronnum-Hansen:2021olh,Liu:2020kpc,Liu:2021wks}.

\section{Conclusions}
In conclusion, the last two decades have witnessed impressive developments in our understanding
of the mathematics of Feynman integrals and how this mathematics plays a role to simplify the calculation
of complicated scattering amplitudes. 
Thanks to these developments, results for very non-trivial scattering amplitudes at two and three loops have been obtained, which were entirely out of reach till few years ago.
These have in turn made it possible to push the boundaries of precision calculations at the LHC to the NNLO
level, reaching the few
percent level precision for many interesting processes and observables. 
A plethora of new possibilities for precision phenomenological studies are opening, and possibly many other 
will open in the next few years, contributing to make the physics program at the LHC a success.

\section*{Acknowledgements}
This work has been supported
by the Excellence Cluster ORIGINS funded by the
Deutsche Forschungsgemeinschaft (DFG, German Research Foundation) under Germany's Excellence Strategy - EXC-2094 - 390783311,
by the ERC Starting Grant 949279 HighPHun
and by the Royal Society grant URF/R1/191125.

\bibliographystyle{JHEP}
\bibliography{biblio.bib}

\end{document}